\documentclass[12pt]{iopart}
\usepackage{iopams}
\usepackage{color}

\begin{document}

\title[A Non-standard Lax Formulation of the HD]{ A non-standard Lax formulation of the Harry Dym hierarchy and its supersymmetric extension}

\author{Kai Tian$^1$, Ziemowit Popowicz$^2$ and Q.~P.~Liu$^1$}

\address{$^1$ Department of Mathematics, China University of Mining
and Technology, Beijing 100083, P. R. China}
\address{$^2$ Institute of Theoretical Physics, University of Wroc{\l}aw, pl. M. Borna 9,50-205, Wroc{\l}aw, Poland}

\eads{\mailto{tiankai@lsec.cc.ac.cn}, \mailto{ziemek@ift.uni.wroc.pl}, \mailto{qpl@cumtb.edu.cn}}

\begin{abstract}
For the Harry Dym hierarchy, a non-standard Lax formulation is deduced from that of Korteweg-de Vries (KdV) equation through
a reciprocal transformation. By supersymmetrizing this Lax operator, a new $N=2$ supersymmetric extension of the Harry Dym hierarchy is constructed, and is further shown to be linked to one of the $N=2$ supersymmetric KdV equations through superconformal transformation. The bosonic limit of this new $N=2$ supersymmetric Harry Dym equation is related to a coupled system of KdV-MKdV equations.
\end{abstract}

\maketitle

\section{Introduction}
The classical integrable systems, alias soliton equations, admit various extensions, among which supersymmetric extensions have been studied extensively in the past three decades. Dealing with supersymmetrization of classical integrable systems, we should distinguish the extended case from the non-extended case. In the non-extended case, one just generalizes the classical theory by pairing classical bosonic fields with some fermionic fields. When all fermionic sectors vanish, the supersymmetric theory degenerates to the known classical one. However in the extended case, in addition to  fermionic fields, new bosonic fields are also introduced into the supersymmetric theory. As a bonus, if we throw fermionic sectors away, certain new classical systems may appear.

As two famous prototypes in the theory of integrable systems, both the Korteweg-de Vries (KdV) and the Harry Dym (HD) equations were successfully supersymmetrized in both  non-extended and  extended cases. Here we only mention some results on $N=2$ supersymmetric KdV and HD equations, which belong to the extended case.

There are three inequivalent $N=2$ supersymmetric KdV equations, which are encoded in the the following  model (denoted by $\mbox{SKdV}_a$)
\begin{equation}\label{kdva}
\Phi_{\tau}=\frac{1}{4}\Big(\Phi_{3y}-3[\Phi(\mathbb{D}_1\mathbb{D}_2\Phi)]_y-\frac{a-1}{2}[\mathbb{D}_1\mathbb{D}_2\Phi^2]_y-3a\Phi^2\Phi_y\Big),
\end{equation}
where $\Phi=\Phi(y,\varrho_1,\varrho_2,\tau)$ is a bosonic super field, and $\mathbb{D}_i=\partial_{\varrho_i}+\varrho_i\partial_x\;(i=1,2)$ denote two super derivatives. Through different approaches \cite{mathieu4,mathieu5}, $\mbox{SKdV}_a$ was shown to be integrable when the parameter $a$ takes $-2$, $1$ or $4$. For these three cases, various integrable features have been revealed, including Lax representations \cite{mathieu4,popo1}, bi-Hamiltonian structures \cite{oevel,kersten} and so on.

By considering the most general $N=2$ Lax operator of \textit{differential} type, two inequivalent $N=2$ supersymmetric HD equations
were constructed \cite{brunelli}, which are respectively given by
\begin{eqnarray}
W_t =& \frac{1}{8}\Big(2W_{3x}W^3-6(\mathcal{D}_1\mathcal{D}_2W_x)(\mathcal{D}_1\mathcal{D}_2W)W^2
-3(\mathcal{D}_2W_{2x})(\mathcal{D}_2W)W^2 \nonumber\\
&\qquad -3(\mathcal{D}_1W_{2x})(\mathcal{D}_1W)W^2\Big),\label{shd4}\\
W_t =& \frac{1}{8}\Big(2W_{3x}W^3-3(\mathcal{D}_2W_{2x})(\mathcal{D}_2W)W^2-3(\mathcal{D}_1W_{2x})(\mathcal{D}_1W)W^2 \nonumber\\
&\qquad +3(\mathcal{D}_2W)(\mathcal{D}_1W)(\mathcal{D}_1\mathcal{D}_2W_x)W\Big),\label{shdm2}
\end{eqnarray}
where $W=W(x,\theta_1,\theta_2,t)$ is a bosonic super field and $\mathcal{D}_i=\partial_{\theta_i}+\theta_i\partial_x\;(i=1,2)$ denote super derivatives. Apart from Lax formulations, few properties have been established  for these two equations until now. It is noticed that the bosonic limits of them were shown to be transformed into previously known integrable systems by reciprocal transformations \cite{sako}.

Very recently, via superconformal transformation both $N=2$ supersymmetric HD equations were related to supersymmetric KdV equations \cite{liu2}, more precisely, the equation \eref{shd4} is changed into $\mbox{SKdV}_4$ by superconformal transformation, while the equation \eref{shdm2} to $\mbox{SKdV}_{-2}$. Hence, the superconformal transformation could be seen as a generalization of reciprocal transformation for $N=2$ supersymmetric systems.

Motivated by these interesting connections and the fact that there are three rather than two $N=2$ supersymmetric KdV equations, we conjecture at least one more $N=2$ supersymmetric HD equation would exist, which is expected to serve as an counterpart of $\mbox{SKdV}_1$. Guided by the fact that $\mbox{SKdV}_1$ has a non-standard Lax formulation \cite{popo1}
\begin{equation*}
\frac{\partial}{\partial \tau}\mathbb{L}_{1}=[(\mathbb{L}_{1}^3)_{\geq 1},\mathbb{L}_{1}],
\end{equation*}
with
\begin{equation}\label{L1}
\mathbb{L}_1=-\partial_y^{-1}\mathbb{D}_1\mathbb{D}_2(\mathbb{D}_1\mathbb{D}_2+\Phi),
\end{equation}
which supplies  a non-standard Lax operator for classical KdV equation, this paper aims at figuring out the \textit{missing} $N=2$ supersymmetric HD equation by supersymmetrizing a suitable non-standard Lax operator of the classical HD hierarchy.

The paper is organized as follows. In section \ref{hdsec}, based on the well-known non-standard Lax KdV hierarchy, a non-standard Lax formulation is constructed for the classical HD hierarchy via reciprocal transformation and gauge transformation. In section \ref{shdsec}, we generalize the non-standard Lax operator to $N=2$ super space, and construct a new $N=2$ supersymmetric HD hierarchy from it. Furthermore, as we anticipated, the Lax operator of our new $N=2$ supersymmetric HD equation is shown to be linked to that of $\mbox{SKdV}_1$ through superconformal transformation. In section \ref{boslim}, the bosonic limit of our new equation is transformed into the bosonic limit of $\mbox{SKdV}_1$ by reciprocal transformation. The last section is devoted to some conclusions.

\section{A non-standard Lax formulation of the HD hierarchy}\label{hdsec}
In the theory of classical integrable systems, reciprocal transformation plays an important role to connect some evolution equations, and is a powerful tool to investigate integrable properties.
For instance, the HD  equation is  connected to the KdV equation  by this  transformation \cite{calogero}.
On the Lax operator level we have two possibilities  to connect the  HD  hierarchy with the  KdV
hierarchy because there are two different Lax operators for the KdV equation:  the so-called standard and non-standard Lax operators.

Let us briefly demonstrate this connection for the standard Lax operator of the KdV equation \cite{kono}.
The KdV hierarchy can be formulated as
\begin{equation}\label{kdvslax}
 \frac{\partial }{\partial \tau_{2n+1}} L_s = [ (L_s^{\frac{2n+1}{2}})_{\geq 0} , L_s],\qquad n=0,1,2,\cdots
\end{equation}
where $L$ is the standard Lax operator, which is given by
\[
L_s=\partial^2_{y} + u,
\]
and the subscript $_{\geq 0}$ denotes the projection to the differential part.
In the case $n=1$, the Lax equation \eref{kdvslax} produces the KdV equation ($\tau \equiv \tau_3$)
\begin{equation}\label{kdv}
 u_{\tau} = u_{3y} + 6uu_y.
\end{equation}
Applying gauge transformation on the operator $L_s$, we obtain
\begin{equation}
 \tilde L_s = e^{\int_{-\infty}^{y} v(z) dz}  L_s e^{-\int_{-\infty}^{y} v(z) dz} = \partial_{y}^2 - 2v\partial_y + u - v_x + v^2.
\end{equation}
Now let $u = v_x - v^2$, then we conclude that the operator $\tilde L_s$ is the Lax operator of the modified KdV equation.
Taking the Liouville transformation $\partial_y= w(y)\partial_x$, the Lax operator $\tilde L_s $ is brought into
\begin{equation}
\hat L_s = w^2\partial_{2x} + (w_y - 2v w)\partial_y,
\end{equation}
assuming that $v=\frac{w_y}{2w}$ the Lax operator $\hat L_s$ is nothing but the standard Lax operator for the HD hierarchy.

On the other hand, the KdV hierarchy can also be formulated as \cite{kono}
\begin{equation}\label{kdvlax}
\frac{\partial}{\partial \tau_{2n+1}}L=[(L^{2n+1}_{\geq 1}),\;L],\qquad n=0,1,2,\cdots
\end{equation}
where $L$ is a non-standard Lax operator, which reads
\[
L=\partial_y^{-1}(\partial_y^2+u).
\]
and subscript $_{\geq 1}$denotes the projection to the part with the powers greater or equal to $\partial_y$.
In the case $n=1$  the Lax equation \eref{kdvlax} implies the KdV equation  \eref{kdv}

Applying gauge transformation on the operator $L$, we have
\begin{eqnarray*}
\tilde{L} &\equiv  \phi^{-1}L\phi = \phi^{-1}\partial_y^{-1}\phi\cdot\phi^{-1}\Big(\partial_y^2+u\Big)\phi\\
& = \phi^{-1}\partial_y^{-1}\phi\cdot\Big(\partial_y^2+2\phi_y\phi^{-1}\partial_y+\phi_{2y}\phi^{-1}+u\Big).
\end{eqnarray*}

Let
\[
\phi=w^{-\frac{1}{2}},\qquad u=\frac{1}{2}w_{2y}w^{-1}-\frac{3}{4}w_y^2w^{-2},
\]
then we have

\[
\tilde{L} = w^{1\over2}\partial_y^{-1}w^{-1\over2}\cdot\Big(\partial_y^2-w_yw^{-1}\partial_y\Big).
\]
It is straightforward to check that the Lax equation
\[
\frac{\partial}{\partial \tau}\tilde{L}=[\tilde{B},\;\tilde{L}] \quad \mbox{with}\quad \tilde{B}=\partial_y^3-\frac{3}{2}w_yw^{-1}\partial_y^2
\]
implies
\begin{equation}\label{sch}
w_\tau=\partial_y\left(w_{2y}-\frac{3}{2}w_y^2w^{-1}\right).
\end{equation}

Since the equation \eref{sch} is a conservation law itself, we introduce a reciprocal transformation as follows
\begin{equation*}
\mathrm{d}x = w\mathrm{d}y+\left(w_{2y}-\frac{3}{2}w_y^2w^{-1}\right)\mathrm{d}\tau,\quad
\mathrm{d}t = \mathrm{d}\tau
\end{equation*}
or equivalently
\begin{equation}\label{recib}
\frac{\partial}{\partial y}=w\frac{\partial}{\partial x},\quad
\frac{\partial}{\partial \tau}=\frac{\partial}{\partial t}+\left(w_{2y}-\frac{3}{2}w_y^2w^{-1}\right)\frac{\partial}{\partial x}
\end{equation}
under which, the equation \eref{sch} is converted into the classical HD equation
\begin{equation}\label{hd}
w_t=w^3w_{3x}.
\end{equation}

Via the same transformation, the operator $\tilde{L}$ is brought to
\begin{equation}\label{nsLax}
\hat{L}=\tilde{L}|_{\partial_y=w\partial_x}=w^{1\over2}\partial_y^{-1}w^{1\over2}\partial_y^2,
\end{equation}
which serves as a non-standard Lax operator for the HD equation \eref{hd}. In fact, the HD equation \eref{hd} is equivalent to
\[
\frac{\partial}{\partial t}\hat{L}=[(\hat{L}^3)_{\geq 2},\;\hat{L}].
\]
and the whole HD hierarchy can be formulated as
\[
\frac{\partial}{\partial t_{2n+1}}\hat{L}=[(\hat{L}^{2n+1})_{\geq 2},\;\hat{L}],\qquad n=0,1,2,\cdots
\]

The nonstandard Lax operator \eref{nsLax}  could be considered as the member of larger class of the  nonstandard HD hierarchy
studied in \cite{kono}. Indeed  this nonstandard HD hierarchy is generated by the following Lax operator
\begin{equation}
 L=w\partial_x + \sum_{k=0}^{\infty} u_k \partial_x^{-k}
\end{equation}
where $w,u_k, k=1,2,\dots$ are  arbitrary functions.  Using the formula
\begin{equation}
 \partial_x^{-1}\cdot f = \sum_{k=0}^{\infty}(-1)^k f_{kx} \partial_x^{-k}
\end{equation}
we can compute all  $u_k$ as  the functions of $w,w_x,w_{xx}\dots$ .

\section{A new $N=2$ supersymmetric HD hierarchy}\label{shdsec}
In this section, we intend to construct a  new $N=2$ HD hierarchy. The idea employed here is to manage to find a proper $N=2$ supersymmetric  analogue of the Lax operator $\hat{L}$ above section.  By trial and error, we find the Lax operator
\[
L=W^{1\over2}\partial_x^{-1}\mathcal{D}_1\mathcal{D}_2W^{1\over2}\mathcal{D}_1\mathcal{D}_2,
\]
where $W=W(x,\theta_1,\theta_2,t)$ is a bosonic super field. The associated Lax equation
\[
\frac{\partial}{\partial t}L=[(L^3)_{\geq 2},\; L]
\]
implies the equation
\begin{eqnarray}
W_t =& -W_{3x}W^3 + \frac{3}{2}(\mathcal{D}_1\mathcal{D}_2W_x)(\mathcal{D}_1\mathcal{D}_2W)W^2 + \frac{3}{2}(\mathcal{D}_2W_{2x})(\mathcal{D}_2W)W^2 \nonumber \\
& + \frac{3}{2}(\mathcal{D}_1W_{2x})(\mathcal{D}_1W)W^2 - \frac{3}{4}(\mathcal{D}_2W)(\mathcal{D}_1W)(\mathcal{D}_1\mathcal{D}_2W_x)W \label{n2hd}.
\end{eqnarray}
which is a new $N=2$ supersymmetric HD equation.
In the remaining part of this section, we will show that the Lax operator $L$ is linked to the Lax operator $\mathbb{L}_1$  \eref{L1} via  a superconformal transformation.

We recall that by a superconformal transformation it means  a super differeomophism transforming the super derivatives covariantly (see \cite{mathieu1} and references there). Let us consider the super diffeomorphism between the super space variables  $(y,\varrho_1,\varrho_2)$ and  $(x,\theta_1,\theta_2)$
\begin{eqnarray}
y \rightarrow  x=x(y,\varrho_1,\varrho_2),\\
\varrho_1 \rightarrow  \theta_1=\theta_1(y,\varrho_1,\varrho_2),\\
\varrho_2 \rightarrow  \theta_2=\theta_2(y,\varrho_1,\varrho_2).
\end{eqnarray}
The associated super derivatives for each super space are respectively denoted by
\[
\mathbb{D}_k=\partial_{\varrho_k}+{\varrho_k}\partial_y,
\qquad\mathcal{D}_k=\partial_{\theta_k}+{\theta_k}\partial_x,\quad
(k=1, 2).
\]

To ensure the super derivatives transforming covariantly, the super diffeomorphism must be constrained by
\begin{equation}\label{eq14}
(\mathbb{D}_1x)  = \theta_1(\mathbb{D}_1\theta_1)+\theta_2(\mathbb{D}_1\theta_2),\quad
(\mathbb{D}_2x)  = \theta_1(\mathbb{D}_2\theta_1)+\theta_2(\mathbb{D}_2\theta_2),
\end{equation}
and
\begin{equation}\label{eq19}
(\mathbb{D}_1\theta_2)  = -(\mathbb{D}_2\theta_1), \quad
(\mathbb{D}_2\theta_2)  = (\mathbb{D}_1\theta_1).
\end{equation}

Under the constraints \eref{eq14} and \eref{eq19}, the super diffeomorphism is a superconformal transformation, which may be formulated by
\numparts
\begin{eqnarray}
\mathcal{D}_1 = K^{-1}\Big((\mathbb{D}_1\theta_1)\mathbb{D}_1+(\mathbb{D}_2\theta_1)\mathbb{D}_2\Big), \label{eq18a}\\
\mathcal{D}_2 = K^{-1}\Big(-(\mathbb{D}_2\theta_1)\mathbb{D}_1+(\mathbb{D}_1\theta_1)\mathbb{D}_2\Big), \label{eq18b}
\end{eqnarray}
\endnumparts
where $K=(\mathbb{D}_1\theta_1)^2+(\mathbb{D}_2\theta_1)^2$.

For the superconformal transformation, the following formulas hold \cite{liu2}
\begin{eqnarray}
\mathcal{D}_1\mathcal{D}_2 = K^{-1}\left[\mathbb{D}_1\mathbb{D}_2+\frac{1}{2}(\mathbb{D}_2\ln K)\mathbb{D}_1-\frac{1}{2}(\mathbb{D}_1\ln K)\mathbb{D}_2\right], \label{eq31}\\
\partial_x = K^{-1}\left[\partial_y-\frac{1}{2}(\mathbb{D}_2\ln K)\mathbb{D}_2-\frac{1}{2}(\mathbb{D}_1\ln
K)\mathbb{D}_1\right]. \label{eq32}
\end{eqnarray}
Based on them, we can prove 
\begin{equation}\label{eq33}
\partial_x^{-1}\mathcal{D}_1\mathcal{D}_2=\partial_y^{-1}\mathbb{D}_1\mathbb{D}_2.
\end{equation}
In fact, \eref{eq31} and \eref{eq32} may be rewritten as
\begin{eqnarray*}
\mathcal{D}_1\mathcal{D}_2 &= K^{-1}\left[\mathbb{D}_1\mathbb{D}_2 + \mathbb{D}_2U\mathbb{D}_1 - U\mathbb{D}_2\mathbb{D}_1 - \mathbb{D}_1U\mathbb{D}_2 + U\mathbb{D}_1\mathbb{D}_2\right]\\
&= K^{-1}\left[\mathbb{D}_1\mathbb{D}_2 + 2U\mathbb{D}_1\mathbb{D}_2 + \mathbb{D}_2U\mathbb{D}_1\partial_y^{-1}\mathbb{D}_2^2 - \mathbb{D}_1U\mathbb{D}_2\partial_y^{-1}\mathbb{D}_1^2\right]\\
& = K^{-1}\left[1 + 2U - \mathbb{D}_2U\partial_y^{-1}\mathbb{D}_2 - \mathbb{D}_1U\partial_y^{-1}\mathbb{D}_1\right]\mathbb{D}_1\mathbb{D}_2,\\
\partial_x &= K^{-1}\left[\partial_y-\mathbb{D}_2U\mathbb{D}_2+U\partial_y-\mathbb{D}_1U\mathbb{D}_1+U\partial_y\right]\\
& = K^{-1}\left[1+2U-\mathbb{D}_2U\partial_y^{-1}\mathbb{D}_2-\mathbb{D}_1U\partial_y^{-1}\mathbb{D}_1\right]\partial_y,
\end{eqnarray*}
where $U=(\ln K)/2$. Now, \eref{eq33} follows from above two equations easily.  We remark that this same equation \eref{eq33} was proved by Mathieu in a different way \cite{mathieu1}.

According to \eref{eq31} and \eref{eq33}, the Lax operator $L$ is transformed to
\begin{equation*}
\tilde{L} = W^{1\over2}\partial_y^{-1}\mathbb{D}_1\mathbb{D}_2W^{1\over2}K^{-1}\left[\mathbb{D}_1\mathbb{D}_2+\frac{1}{2}(\mathbb{D}_2\ln K)\mathbb{D}_1-\frac{1}{2}(\mathbb{D}_1\ln K)\mathbb{D}_2\right].
\end{equation*}
Applying gauge transformation on it and let $K=W$, we have
\begin{eqnarray*}
W^{-\frac{1}{2}}\tilde{L}W^{1\over2} &= \partial_y^{-1}\mathbb{D}_1\mathbb{D}_2W^{-\frac{1}{2}}\left[\mathbb{D}_1\mathbb{D}_2+\frac{1}{2}(\mathbb{D}_2\ln W)\mathbb{D}_1-\frac{1}{2}(\mathbb{D}_1\ln W)\mathbb{D}_2\right]W^{1\over2}\\
&= \partial_y^{-1}\mathbb{D}_1\mathbb{D}_2\left[\mathbb{D}_1\mathbb{D}_2+\frac{1}{2}\frac{(\mathbb{D}_1\mathbb{D}_2W)}{W} + \frac{3}{4}\frac{(\mathbb{D}_2W)(\mathbb{D}_1W)}{W^2}\right]
\end{eqnarray*}
Assuming
\begin{eqnarray*}
\Phi &= \frac{1}{2}\frac{(\mathbb{D}_1\mathbb{D}_2W)}{W} + \frac{3}{4}\frac{(\mathbb{D}_2W)(\mathbb{D}_1W)}{W^2}\\
&= \mathbb{D}_1\mathbb{D}_2\left(\frac{1}{2}\ln W\right)+\mathbb{D}_2\left(\frac{1}{2}\ln W\right)\mathbb{D}_1\left(\frac{1}{2}\log W\right)
\end{eqnarray*}
we arrive at the Lax operator $\mathbb{L}_1$ of the $\mbox{SKdV}_1$ equation.

In terms of the new variable $U=(\ln W)/2$, the operator $\tilde{L}$ reads as
\[
\tilde{L}=e^U\partial_y^{-1}\mathbb{D}_1\mathbb{D}_2e^{-U}\Big[\mathbb{D}_1\mathbb{D}_2 + (\mathbb{D}_2U)\mathbb{D}_1 - (\mathbb{D}_1U)\mathbb{D}_2\Big],
\]
which generates a modified equation of $\mbox{SKdV}_1$. Indeed, the Lax equation
\[
\frac{\partial}{\partial \tau}\tilde{L}=[(\tilde{L}^3)_{\geq 1},\; \tilde{L}]
\]
yields the equation
\begin{eqnarray*}
U_\tau =& -U_{3y} + 2U_y^3 + 3(\mathbb{D}_1\mathbb{D}_2U_y)(\mathbb{D}_1\mathbb{D}_2U) + 3(\mathbb{D}_1\mathbb{D}_2U)^2U_y \\
& + 3(\mathbb{D}_2U_y)(\mathbb{D}_2U)U_y + 3(\mathbb{D}_1U_y)(\mathbb{D}_1U)U_y + 3(\mathbb{D}_2U_y)(\mathbb{D}_1U)(\mathbb{D}_1\mathbb{D}_2U) \\
& + 3(\mathbb{D}_2U)(\mathbb{D}_1U_y)(\mathbb{D}_1\mathbb{D}_2U) + 3(\mathbb{D}_2U)(\mathbb{D}_1U)(\mathbb{D}_1\mathbb{D}_2U_y).
\end{eqnarray*}
Thus, we have proved that the new supersymmetric HD equation \eref{n2hd} is transformed to $\mbox{SKdV}_1$ by means of superconformal transformation.

\section{Bosonic limit of the new $N=2$ supersymmetric HD equation}\label{boslim}
In this section, we consider the bosonic limit of the new $N=2$ supersymmetric HD equation, and relate it to the bosonic limit of the  $\mbox{SKdV}_1$ equation  through a series of transformations, among which a reciprocal transformation is the key ingredient.

For our new $N=2$ supersymmetric HD equation
\begin{eqnarray*}
W_t =& -W_{3x}W^3 + \frac{3}{2}(\mathcal{D}_1\mathcal{D}_2W_x)(\mathcal{D}_1\mathcal{D}_2W)W^2 + \frac{3}{2}(\mathcal{D}_2W_{2x})(\mathcal{D}_2W)W^2 \\
& + \frac{3}{2}(\mathcal{D}_1W_{2x})(\mathcal{D}_1W)W^2 - \frac{3}{4}(\mathcal{D}_2W)(\mathcal{D}_1W)(\mathcal{D}_1\mathcal{D}_2W_x)W,
\end{eqnarray*}
by setting
\[
W(x,\theta_1,\theta_2,t)=w_0(x,t)+\theta_2\theta_1 w_1(x,t),
\]
we obtain its bosonic limit
\numparts
\begin{eqnarray}
w_{0,t} =& -w_{0,3x}w_0^3 + \frac{3}{2}w_{1,x}w_1w_0^2, \label{nhdba}\\
w_{1,t} =& -w_{1,3x}w_0^3 - 3w_{1,2x}w_{0,x}w_0^2 - \frac{3}{2}w_{1,x}w_{0,2x}w_0^2 - \frac{3}{4}w_{1,x}w_{0,x}^2w_0 \nonumber\\
& + \frac{9}{4}w_{1,x}w_1^2w_0 - \frac{3}{2}w_1w_{0,3x}w_0^2. \label{nhdbb}
\end{eqnarray}
\endnumparts
It is obvious that a further reduction $w_1=0$ results in the Harry Dym equation.

The coupled system \eref{nhdba}\eref{nhdbb} admits a conservation law
\[
\frac{\partial}{\partial t}\Big(w_0^{-1}\Big) = \frac{\partial}{\partial x}\left(w_{0,2x}w_0-\frac{1}{2}w_{0,x}^2-\frac{3}{4}w_1^2\right).
\]
Based on this conservation law, new independent variables could be introduced as
\begin{equation*}
\mathrm{d}y = w_0^{-1}\mathrm{d}x + \left(w_{0,2x}w_0-\frac{1}{2}w_{0,x}^2-\frac{3}{4}w_1^2\right)\mathrm{d}t, \qquad \mathrm{d}\tau =\mathrm{d}t.
\end{equation*}
In terms of new variables $(y,\tau)$, the evolution of $(w_0,w_1)$ is governed by
\numparts
\begin{eqnarray}
w_{0,\tau} &= -w_{0,3y} + 3w_{0,2y}w_{0,y}w_0^{-1}-\frac{3}{2}w_{0,y}^3w_0^{-2} + \frac{3}{2}w_0w_{1,y}w_1 + \frac{3}{4}w_{0,y}w_1^2, \label{cscha} \\
w_{1,\tau} &= -w_{1,3y} + \frac{9}{4}w_{1,y}w_{0,y}^2w_0^{-2} - \frac{3}{2}w_{1,y}w_{0,2y}w_0^{-1} + 3w_{1,3y}w_1^2 \nonumber\\
& - \frac{3}{2}w_1w_{0,3y}w_0^{-1} + 6w_1w_{0,2y}w_{0,y}w_0^{-2} - \frac{9}{2}w_1w_{0,y}^3w_0^{-3}.\label{cschb}
\end{eqnarray}
\endnumparts

Assuming 
\[
u_0=-(\ln w_0)_y, \qquad u_1=w_1,
\]
the system \eref{cscha}\eref{cschb} is converted into a coupled system of modified KdV equations
\numparts
\begin{eqnarray}
u_{0,\tau} = \left(-u_{0,2y} + \frac{1}{2}u_0^3 - \frac{3}{2}u_{1,y}u_1 + \frac{3}{4}u_0u_1^2\right)_y,\\
u_{1,\tau} = \left(-u_{1,2y} + u_1^3 + \frac{3}{4}u_0^2u_1 + \frac{3}{2}u_{0,y}u_1\right)_y.
\end{eqnarray}
\endnumparts
Finally, with the help of the following Miura-type transformation
\[
v_0 = \frac{1}{2}u_{0,y} + \frac{1}{4}u_0^2 + \frac{1}{4}u_1^2 \qquad v_1 =\frac{1}{2}u_1,
\]
we arrive at the bosonic limit of $\mbox{SKdV}_1$
\numparts
\begin{eqnarray}
v_{0,\tau} = \Big(-v_{0,2y} + 3v_0^2 + 3v_0v_1^2 - 3v_{1,2y}v_1\Big)_y, \label{ffeqa}\\
v_{1,\tau} = \Big(-v_{1,2y} + v_1^3 + 3v_0v_1\Big)_y, \label{ffeqb}
\end{eqnarray}
\endnumparts
which can be obtained by substituting $\Phi=v_0+\theta_2\theta_1 v_2$ into $\mbox{SKdV}_1$ {\cite{popo1}. The integrability of the system \eref{ffeqa}\eref{ffeqb} was also confirmed by the existence of recursion operator and infinitely many symmetries \cite{kers}.

\section{Conclusions}
By studying the non-standard Lax representation for the HD hierarchy, we have successfully constructed a new $N=2$ supersymmtric HD hierarchy, thus there are three integrable HD type hierarchies, which are comparable with the $N=2$ supersymmetric KdV hierarchies. We further show that this new HD hierarchy, like the other two, is also connected with one of the supersymmetric KdV hierarchies through supercomformal transformation.  The further properties of this new system deserves further investigation which may be done in future work.

\bigskip
\noindent
\textbf{Acknowledgement} This work is supported by the National
Natural Science Foundation of China (grant numbers: 10671206,
10731080, 10971222) and the Fundamental Research Funds for Central
Universities.

\end{document}